\documentclass[num-refs]{wiley-article}

\usepackage{mdsymbol}

\usepackage{siunitx}
\usepackage[title]{appendix}

\papertype{Original Article}
\paperfield{Journal Section}
\newtheorem{prop}{Proposition}

\title{Properties of a confirmatory two-stage adaptive procedure for assessing average bioequivalence}

\author[1\authfn{1}]{Marie Louise {\O}sterdal, MSc.}
\author[1\authfn{1}]{Kyle Raymond, PhD}
\author[1,2\authfn{1}]{Christian Pipper, PhD}

\contrib[\authfn{1}]{Equally contributing authors.}

\affil[1]{Dept of Biostatistics and Pharmacoepidemiology, LEO Pharma A/S}
\affil[2]{Dept. of Public Health, University of Southern Denmark}

\corraddress{Christian Pipper PhD, Biostatistics 1, LEO Pharma A/S, 2750 Ballerup, Denmark}
\corremail{TPBDK@leo-pharma.com}

\runningauthor{{\O}sterdal et al.}

\begin{document}

\begin{frontmatter}
\maketitle

\begin{abstract}
We investigate a confirmatory two stage adaptive procedure for assessing average bioequivalence and provide some insights to its theoretical properties. Effectively, we perform \textbf{T}wo \textbf{O}ne-\textbf{S}ided \textbf{T}ests (TOST) to reach overall decision about each of the two traditional null-hypotheses involved in declaring average bioequivalence. The tests are performed as combination tests separately for each hypothesis based on the corresponding pair of stagewise p-values.  Features of the procedure include a built in futility, sample size reassessment, and the ability to simultaneously assess average bioequivalence with respect to multiple endpoints while controlling the familywise error rate. To facilitate inference at the end of a trial we also derive overall confidence limits that match the decision reached on each one sided hypothesis and provide theory ensuring their appropriateness. The performance is assessed by simulation in the context of planning a study to compare two different administrations of a biologic treatment of atopic dermatitis.
% Please include a maximum of seven keywords
\keywords{\emph{Adaptive design}, \emph{Bioequvalence},\emph{Combination tests}, \emph{Multiple endpoints}, \emph{Overall confidence limits}}
\end{abstract}
\end{frontmatter}

\section{Introduction}

A two-stage design for a bioequivalence assessment is a valuable tool when little is known about the variability of the measured PK parameters and the true difference between the two formulations for which bioequivalence must be demonstrated. 
There is an extensive litterature on the use of two-stage group-sequential or adaptive designs in such a setting, but rather few examples of the use of these methods in practice \cite{kaza}. Specifically, confirmatory procedures that apply an adaptive sample size re-estimation after stage 1 have been suggested in for instance \cite{maurer} and in practice several flexible software platforms enabling implementation of adaptive trials in the context of bioequivalence assessment are already established. However, even though all the tools to perform an adaptive bioequivalence trial are readily available a number of properties ensuring proper performance are still unclear. To fill this gap we provide some theoretical properties in the context of:
\begin{itemize}
\item[1)] Implementation of a binding futility bound, aiming to prevent fruitless trials from continuing after the first stage, 

\item[2)] Producing final confidence intervals that combine information from both stages and are consistent with the decision making,

\item[3)] The ability to efficiently handle multiple endpoints in the context of assessing simultaneous average bioequivalence. 
\end{itemize}
Specifically, this paper considers a confirmatory adaptive \textbf{T}wo \textbf{O}ne-\textbf{S}ided \textbf{T}ests (TOST) procedure with a built in  futility bound, adapted to handle simultaneous bioequivalence of multiple pharmacokinetics endpoints, and accompanied by an overall confidence region that is ensured to support the final bioequivalence decision.

Following the above strategy a number of desirable properties are not automatically ensured and need further scrutiny.  Firstly, it is not per default ensured that the upper limit of the suggested overall confidence interval is in fact higher than the lower limit. In this paper we show that under some mild assumptions this does in fact hold true. 

Secondly, the traditional approach for assessing the simultaneous bioequivalence by using the maximal p-value to test the union of null-hypotheses against the intersection of the alternative hypotheses does not utilize the correlation between p-values and may therefore provide conservative control of the type 1 error. We provide an alternative procedure for simultaneously assessing bioequivalence and show that this procedure not only provides type 1 error control but is also more powerful than the traditional Intersection Union approach \cite{berger}.

%Also, it has been proposed [ref: Maurer 2018] to use a pre-specified geometric mean ratio in the sample size re-calculation step, whereas we explore an apprach where the stage 1 estimate is used. Finally, we explore the implication of varying choices of the weights used in the maximum combination test suggested by [ref: Maurer 2018]. 

In a traditional fixed design setting the two one-sided test (TOST) procedure for assessing bioequivalence considers testing the null hypothesis: 
$$
H_{0}:\:\:|\theta|>\Delta,
$$
where $\theta$ denotes the true value of the targeted population parameter. Due to the make-up of $H_{0}$ as the union of the two one-sided null-hypotheses $H_{0}^{-}:\:\: \theta<-\Delta$ and $H_{0}^{+}:\:\: -\theta<-\Delta$  a decision to reject $H_{0}$ is equivalent to rejecting both $H_{0}^{-}$ and $H_{0}^{+}$.  Accordingly $H_{0}$ is typically evaluated by considering two one-sided tests (TOST) for evaluating    $H_{0}^{-}$ and $H_{0}^{+}$, respectively. The two test statistics used for this evaluation may typically  be represented (at least asymptotically) as follows:
$$
Z_{-}=\varepsilon(\theta)+\frac{\theta+\Delta}{\sigma_{n}} ,\:\:\:\: Z_{+}=-\varepsilon(\theta)+\frac{-\theta+\Delta}{\sigma_{n}}
$$
where $\varepsilon(\theta)=\frac{\hat{\theta}-\theta}{\sigma_{n}}\sim N(0,1)$ and $\sigma_{n}=\sigma/\sqrt{n}$ with $\sigma$ denoting an appropriate dispersion parameter and $n$ denoting the sample size. Large values of the tests are considered critical and the resulting p-values are given as

$$
p_{-}=1-\Phi(Z_{-}),\:\:\:\: p_{+}=1-\Phi(Z_{+}),
$$
where $\Phi$ denotes the cumulative distribution function of the standard normal distribution. 
The null hypothesis of non-bioequivalence is then rejected at significance level $\alpha$ if both tests are rejected at level $\alpha$, that is, if 
$$
\max\{p_{-},p_{+}\}<\alpha.
$$
In this classical setting a two-sided confidence interval with confidence level $1-2\alpha$ is most naturally constructed by, based on the property $\frac{\hat{\theta}-\theta}{\sigma_{n}}\sim N(0,1)$, reverting the standard normal area $[\Phi^{-1}(\alpha);\Phi^{-1}(1-\alpha)]$ to an interval $[l;u]$ in which we are $1-2\alpha$ confident to find the true value $\theta$. 

An alternative way of deriving the same confidence interval in this setting is based on the uniform distribution of the shifted p-values:   
$$
p_{-}^{\theta}=1-\Phi(\varepsilon(\theta)),\:\:\: p_{+}^{\theta}=1-\Phi(-\varepsilon(\theta)).
$$ 
With this approach lower bound is now found as the smallest $\delta$ such that:

$$
p_{-}^{\delta}\geq \alpha
$$
yielding:
$$
l=\hat{\theta}-\Phi^{-1}(1-\alpha)\sigma_{n}.
$$
Similarly an upper bound is found as $-\delta$  with the smallest $\delta$ such that:

$$
p_{+}^{\delta}\geq\alpha
$$
yielding:
$$
u=\hat{\theta}+\Phi^{-1}(1-\alpha)\sigma_{n}.
$$

The p-value based approach outlined above becomes particularly attractive in the adaptive setting we consider, where we only have the uniform distribution of the shifted overall p-values from a combination test to rely on \cite{bran02}. Moreover, by construction, the p-value based confidence interval is aligned to the decision regarding  $H_{0}^{-}$ and $H_{0}^{+}$, that is,   $l<-\Delta$ if and only if $p_{-}=p_{-}^{-\Delta}\geq \alpha/2$  and $u>\Delta$ if and only if $p_{+}=p_{+}^{-\Delta}\geq\alpha/2$.

In Section \ref{adaptost} we describe the two stage confirmatory TOST procedure to test $H_{0}^{-}$ and $H_{0}^{+}$ based on a general combination test procedure. The resulting overall shifted p-values from that procedure are then used to develop an overall $1-2\alpha$ confidence interval for the target parameter $\theta$ that aligns with the final decision about rejection/acceptance of $H_{0}^{-}$ and $H_{0}^{+}$. The resulting procedure is then extended to handle two target parameters simultaneously in Section \ref{multend}. Next, in Section \ref{samplesize}, we outline a sample size re-estimation procedure intended to ensure a given overall power. All developments are then combined in Section \ref{Conc ex} where we present simulation results to support the planning of a bioequivalence study comparing two different administrations of a biologic treatment of moderate to severe atopic dermatitis. Finally, Section \ref{discuss} contains a discussion of the developed methods.

\section{An adaptive TOST procedure}\label{adaptost}

To fix notation, let the stage wise test statistics for stage $j=1,2$ be defined by 
$$
Z_{j-}=\varepsilon_{j}(\theta)+\frac{\theta+\Delta}{\sigma_{nj}} ,\:\:\:\: Z_{j+}=-\varepsilon_{j}(\theta)+\frac{-\theta+\Delta}{\sigma_{nj}}
$$
where $\varepsilon_{1}(\theta)\upvDash \varepsilon_{2}(\theta)$ and $\varepsilon_{j}(\theta)=\frac{\hat{\theta}_{j}-\theta}{\sigma_{nj}}\sim N(0,1)$ and $\sigma_{nj}=\sigma/\sqrt{n_{j}}$ with $\sigma$ denoting an appropricate dispersion parameter and $n_{j}$ denoting the sample size $j$. Large values of the tests are considered critical and the resulting stagewise p-values are given as: 
$$
p_{j-}=1-\Phi(Z_{j-}),\:\:\:\: p_{j+}=1-\Phi(Z_{j+}).
$$

For an efficacy bound $\alpha_{1}$, futility bound $\alpha_{0}$, and a corresponding combination function $C$ the overall p-values are given by $Q(p_{1-},p_{2-},\alpha1,\alpha_{0})$ and $Q(p_{1+},p_{2+},\alpha_{1},\alpha_{0})$ where

\begin{align*}
Q(p_{1},p_{2},\alpha_{1},\alpha_{0})&=p_{1}(I(p_{1}\leq\alpha_{1})+I(p_{1}<\alpha_{0}))+I(\alpha_{0}>p_{1}>\alpha_{1})[\alpha_{1}+\int_{\alpha_{1}}^{\alpha_{0}}\int_{0}^{1}I\{C(x,y)\leq C(p_{1},p_{2})\}dydx]\\
&=p_{1}I((p_{1}<\alpha_{1})+I(p_{1}>\alpha_{0}))+I(\alpha_{0}>p_{1}\geq\alpha_{1})(\alpha_{1}+P(\alpha_{0}>X\geq\alpha_{1},C(X,Y)\leq C(p_{1},p_{2}))
\end{align*}
for $X$ and $Y$ independent and uniformly distributed. 
The null hypothesis of non-bioequivalence is then rejected at significance level $\alpha$ according to the union-intersection principle \cite{berger}, that is, if 
$$
\max\{Q(p_{1-}.p_{2-},\alpha_{1},\alpha_{0}),Q(p_{1+},p_{2+},\alpha_{1},\alpha_{0})\}<\alpha
$$

Notice that this decision rule is not similar to the one described in \cite{maurer}. There, if one proceeds to stage 2, both hypotheses $H_{0}^{+}$ and $H_{0}^{-}$ are re-evaluated even if one of them  is formally  rejected at stage 1. In the context of \cite{maurer}, where stopping for futility is not built into the procedure ($\alpha_{0}=1$), such a decision strategy will control the type I error.  However,  in the setup we describe in this paper, where stopping for futility is built into the decision procedure ($\alpha_{0}<1$), re-evaluating a null-hypothesis at stage 2  that has already been accepted at stage 1 would clearly jeopardize type 1 error control.  In terms of power our suggestion also has the advantage that if we reject one of the one-sided hypotheses already at stage 1 but still proceed to stage 2, we don't have to reject that same null-hypothesis again after stage 2.   

To provide a two-sided confidence interval with confidence level $1-2\alpha$ we consider the following shifted  stagewise p-values 

$$
p_{j-}^{\delta}=1-\Phi(\varepsilon_{j}(\delta)),\:\:\: p_{j+}^{\delta}=1-\Phi(-\varepsilon_{j}(-\delta))
$$ 
Testing the hypotheses:
$$
H_{0}^{\delta-}:\:\: \theta\leq\delta,\:\:\: H_{0}^{\delta+}:\:\:-\theta\leq\delta.  
$$
Furthermore define
$$
\alpha_{1}^{\delta}=1-\Phi(\Phi^{-1}(1-\alpha_{1})-\frac{\delta+\Delta}{\sigma_{1n}}),\:\:\alpha_{0}^{\delta}=1-\Phi(\Phi^{-1}(1-\alpha_{0})-\frac{\Delta+\delta}{\sigma_{1n}})
$$
Finally define the overall shifted p-values as
$$
p_{-}^{\delta}=Q(p_{1-}^{\delta},p_{2-}^{\delta},\alpha_{1}^{\delta},\alpha_{0}^{\delta});\:\:p_{+}^{\delta}=Q(p_{1+}^{\delta},p_{2+}^{\delta},\alpha_{1}^{\delta},\alpha_{0}^{\delta})
$$
As before we identify the lower bound $l$ as the smallest $\delta$ such that:

\begin{equation}\label{lower}
p_{-}^{\delta}\geq \alpha
\end{equation}
and the  upper bound $u$ is identified as  $-\delta$ with the smallest $\delta$ such that:

\begin{equation}\label{upper}
p_{+}^{\delta}\geq\alpha
\end{equation}

Note that the above definition of the two sided overall confidence interval is aligned to the actual declaration of non-equivalence in the following manner:

$$
p_{j-}=p_{j-}^{-\Delta},\:\:p_{j+}=p_{j+}^{-\Delta}
$$
so that non equivalence is accepted at level $\alpha/2$ exactly if either $-\Delta$ or $\Delta$ or both belong to the constructed two-sided $1-\alpha$ confidence interval.\\\\In a trial with a large enough stage 1 sample we also  have that $l<u$. This is summarized in Proposition \ref{thm1} below the proof of which is presented in Appendix A.1.  
\begin{prop}\label{thm1}
For any $\alpha_{0}\leq0.5$, for $\alpha_{1}<\alpha$ such that $2Z_{1-\alpha}-Z_{1-\alpha_{1}}>0$ , and for $\frac{\Delta}{\sigma_{1n}}>Z_{1-\alpha_{1}}-Z_{1-\alpha}$ it holds that $l<u$. \end{prop}

We note that the  assumption $2\cdot Z_{1-\alpha}-Z_{1-\alpha_{1}}>0$ is more of a technical issue than a practical restriction on $\alpha_{1}$. For instance , with $\alpha=0.05$ this dictates that $\alpha_{1}>0.0005$. 

Also, and perhaps more crucial, note that the formal decision procedure prompts that you may reach a decision not in favor of bioequivalence or partially in favor of bioequivalence at stage 1 but still go on to stage 2.  For instance you could decide in favor of not  rejecting or rejecting  $H_{0}^{-}$ at stage 1, but still go on to stage 2 to decide whether to reject $H_{0}^{+}$. In this case the adaptive design, if followed as described,  does not use stage 2 data for a re-evaluation of  $H_{0}^{-}$.  
\\\\One could rightfully argue that the strategy of continuing to stage 2 when you have already given up on declaring bioequivalence is fruitless. However, on the other hand, it may still be of value to learn if either non-inferiority or non-superiority may be declared  in the absence of the other, for instance if you wanted to adjust the dosing of your test formulation or test device. 
To this end also notice that if stage 1 is designed to be large enough to ensure that $p_{1-}>\alpha_{0}$ dictates $p_{1+}<\alpha_{1}$ or vice versa we would always stop at stage 1 if we give up on declaring bioequivalence at that stage. This would be the case if we could ensure that $\frac{2\Delta}{\sigma_{1n}}>Z_{1-\alpha_{1}}+Z_{1-\alpha_{0}}$.

Finally, note that according to Proposition A.1. in \cite{bran02}  the shifted p-values are uniformly distributed for $\delta=\theta$. Consequently $\theta$ will be below $l$  with probability $\alpha$ and above $u$ with probability $\alpha$. In the above exposition this of course relies on us knowing the true value of the  dispersion parameter $\sigma$. This would never be the case in practice and we would replace $Z_{j-}$ and $Z_{j+}$ by the corresponding standard $t$-statistics:
$$
T_{j-}=\frac{\sqrt{ n_{j}}(\hat{\theta}_{j}+\Delta)}{\hat{\sigma}_{j}} ,\:\:\:\: T_{j+}=\frac{\sqrt{n_{j}}(-\hat{\theta}_{j}+\Delta)}{\hat{\sigma}_{j}}, 
$$
where $\hat{\sigma}_{j}$ is an estimator of $\sigma$ based on $j$th stage data. This has the consequence that when applied as is the adaptive TOST only ensures type 1 error control and correct coverage of confidence intervals asymptotically. It is possible to improve small sample behavior by making assumptions about the exact simultaneous distribution of $(\hat{\theta}_{j},\hat{\sigma}_{j})$, but we do not pursue this further here. The performance of the proposed adaptive TOST procedure based on standard $t$-statistics is evaluated by simulation in Section \ref{Conc ex}.

\section{Multiple endpoints}\label{multend}
In a pk bioequivalence trial focus is predominantly on establishing equivalence with respect to both log of the  maximal concentration $\log(C_{max})$ and log of the area under the concentration curve $\log(AUC)$ as required by regulatory authorities \cite{pkbioqFDA}.  The traditional approach is to power the study to one of them \cite{maurer}. The rationale for such an approach is that typically the correlation between $\log(C_{max})$ and $\log(AUC)$ will be very high and also the target parameter $\theta^{(1)}$ for $\log(C_{max})$ will be approximately equal to the the target parameter $\theta^{(2)}$ for $\log(AUC)$. As a consequence any assessment with respect to bioequivalence based on $\log(C_{max})$ is expected to be mirrored in a similar assessment based on $\log(AUC)$. Clearly such a strategy is vulnerable even if the rationale is sound due to formal lack of type 1 error control. An excellent and in detail discussion of these challenges is available in \cite{pallmann}.    

We propose a more direct approach based on assessing the target parameters $\min(\theta^{(1)},\theta^{(2)})$ and $\max(\theta^{(1)},\theta^{(2)})$. Specifically we propose to test the following hypotheses: 

\begin{align*}
&H_{0}^{1-}\cup H_{0}^{2-} : \:\:\min(\theta^{(1)},\theta^{(2)})<-\Delta,\\
&H_{0}^{1+}\cup H_{0}^{2+}: \:\:\max(\theta^{(1)},\theta^{(2)})>\Delta
\end{align*}

where $H_{0}^{k-}$ and $H_{0}^{k+}$ denote  $k$th endpoint hypotheses in scope, that is,  

\begin{align*}
&H_{0}^{k-}: \:\:\theta^{(k)}<-\Delta,\\
&H_{0}^{k+}: \:\:\theta^{(k)}>\Delta.
\end{align*}

In order to do so define the $j$th stage estimator of $\min(\theta^{(1)},\theta^{(2)})$ and $\max(\theta^{(1)},\theta^{(2)})$ as:

\begin{align*}
 &\hat{\theta}_{j}^{min}= A_{j}\hat{\theta}_{j}^{(1)}+(1-A_{j})\hat{\theta}_{j}^{(2)},\\
 &\hat{\theta}_{j}^{max}= A_{j}\hat{\theta}_{j}^{(2)}+(1-A_{j})\hat{\theta}_{j}^{(1)},
\end{align*}
where $A_j$ equals one if $\hat{\theta}_{j}^{(1)}\leq \hat{\theta}_{j}^{(2)}$ and zero otherwise. Moreover define:

\begin{align*}
 &\sigma_{jn}^{min}= A_{j}\sigma_{jn}^{(1)}+(1-A_{j})\sigma_{jn}^{(2)},\\
 &\sigma_{jn}^{max}= A_{j}\sigma_{jn}^{(2)}+(1-A_{j})\sigma_{jn}^{(1)}.\\
\end{align*}

Finally define the stagewise tests and p-values as: 

\begin{align*}
 &Z_{j}^{min}=\frac{\hat{\theta}_{j}^{min}+\Delta}{\sigma_{jn}^{min}}=A_{j}Z^{(1)}_{j-}+(1-A_{j})Z^{(2)}_{j-},\\
 &Z_{j}^{max}=\frac{-\hat{\theta}_{j}^{max}+\Delta}{\sigma_{jn}^{max}}=A_{j}Z^{(2)}_{j+}+(1-A_{j})Z^{(1)}_{j+},\\
 &p_{j-}^{min}=1-\Phi(Z_{j}^{min})=A_{j}p_{j-}^{(1)}+(1-A_{j})p_{j-}^{(2)},\\
 &p_{j+}^{max}=1-\Phi(Z_{j}^{max})=A_{j}p_{j+}^{(2)}+(1-A_{j})p_{j+}^{(1)},
\end{align*}
where $Z_{j-}^{(k)},Z_{j+}^{(k)}, p_{j-}^{(k)},p_{j+}^{(k)}$ denote previously defined stagewise test statistics and p-values for the $k$th endpoint.

With these definitions $H_{0}^{1-}\cup H_{0}^{2-}$ is assessed as in section \ref{adaptost} using the stagewise p-values $p_{j-}^{min}$ and  likewise $H_{0}^{1+}\cup H_{0}^{2+}$ is assessed using the stagewise p-values $p_{j+}^{max}$. The overall p-values of these assessments may be calculated as $Q(p_{1}^{min},p_{2}^{min}, \alpha_{1},\alpha_{0})$ and $Q(p_{1}^{max},p_{2}^{max}, \alpha_{1},\alpha_{0})$, respectively. Also mimicking the developments in section \ref{adaptost} the shifted stagewise p-values may be defined as:  

\begin{align*}
    &p_{j-}^{\delta, min}=1-\Phi(\varepsilon^{min}_{j}(\delta)),\\
    &p_{j+}^{\delta, max}=1-\Phi(-\varepsilon^{max}_{j}(-\delta)),
\end{align*}
where 
\begin{align*}
    &\varepsilon_{j}^{min}(\delta)=\frac{\hat{\theta}_{j}^{min}-\delta}{\sigma_{nj}^{min}},\\
    &\varepsilon_{j}^{max}(\delta)=\frac{\hat{\theta}_{j}^{max}-\delta}{\sigma_{nj}^{max}}.
\end{align*}
From the shifted p-values a $1-\alpha$ lower bound $l^{min}$ for $\min(\theta^{(1)},\theta^{(2)})$  as well as a $1-\alpha$ upper bound $u^{max}$ for $\max(\theta^{(1)},\theta^{(2)})$ may be calculated as in Section \ref{adaptost}. Below we present a result showing that under similar assumptions as those imposed in Proposition \ref{thm1} the above procedure leads to a meaningful confidence interval. We first define the following quantities used in the assumptions
\begin{align*}
&\omega=\sigma_{1n}^{min}/\sigma_{1n}^{max},\\
&\kappa=Z_{1-\alpha_{1}}-\frac{2\Delta}{\sigma_{in}^{min}+\sigma_{in}^{max}}.
\end{align*}

\begin{prop}\label{thm1a}
For any $\alpha_{0}\leq0.5$ assume that $\alpha_{1}<\alpha$ is such that $\min\{1+\omega,1+1/\omega\}Z_{1-\alpha}-Z_{1-\alpha_{1}}>0$. Also assume that:
\begin{equation}\label{assump1}
\Phi(\frac{-\omega\kappa+\sqrt{\kappa^2+2\log(w)\frac{w-1}{w+1}}}{w-1})+\Phi(\frac{\kappa-\omega\sqrt{\kappa^2+2\log(w)\frac{w-1}{w+1}}}{w-1})>2\alpha.
\end{equation}
Then it holds that $l^{min}<u^{max}$.
\\\\Furthermore a sufficient condition for (\ref{assump1}) to hold is that
$$
\frac{-\kappa-\sqrt{\kappa^{2}+2\log(w)\frac{w-1}{w+1}}}{2}>-Z_{1-\alpha}.
$$
\end{prop}
Note that compared to Proposition \ref{thm1} the assumption $\min\{1+\omega,1+1/\omega\}Z_{1-\alpha}-Z_{1-\alpha_{1}}>0$ is stricter than the assumption $2Z_{1-\alpha}-Z_{1-\alpha_{1}}>0$. It  requires that the precisions with which we estimate the two endpoint parameters are not to different in the sense that $\frac{Z_{1-\alpha_{1}}-Z_{1-\alpha}}{Z_{1-\alpha}}<\sigma_{1n}^{min}/\sigma_{1n}^{max}<\frac{Z_{1-\alpha}}{Z_{1-\alpha_{1}}-Z_{1-\alpha}}$. When $\sigma_{1n}^{min}=\sigma_{1n}^{max}$ corresponding to $\omega=1$ the first assumption of Proposition \ref{thm1a} reduces to the assumption $2Z_{1-\alpha}-Z_{1-\alpha_{1}}>0$ in Proposition \ref{thm1}. Moreover, both (\ref{assump1}) and the sufficient condition reduce to the assumption.  $\frac{\Delta}{\sigma_{1n}}>Z_{1-\alpha_{1}}-Z_{1-\alpha}$ in Proposition \ref{thm1}. We give a formal proof of Proposition \ref{thm1a} in Appendix A.2.

For the above procedure  to remain a valid procedure in terms of controlling the type 1 error and to ensure coverage of the proposed confidence limits we need to show that the stagewise p-values defined above are asymptotically p-clud \cite{wassmer}. This result is summarized in Proposition \ref{thm2} below. A formal proof of Proposition \ref{thm2} is presented in Appendix A.3.

\begin{prop}\label{thm2} 
If $H_{0}^{1-}\cup H_{0}^{2-}$ is true then for any $0\leq\alpha\leq1$:
$$
    \lim_{n_{j}\rightarrow\infty}P(p_{j-}^{min}\leq\alpha)\leq\alpha.
$$
If $H_{0}^{1+}\cup H_{0}^{2+}$ is true then for any $0\leq\alpha\leq1$ :
$$
    \lim_{n_{j}\rightarrow\infty}P(p_{j+}^{max}\leq\alpha)\leq\alpha.
$$

\end{prop}

  For a traditional approach we note that the adaptive procedure developed in Section \ref{adaptost} may be applied marginally to  assess $H_{0}^{k-}$ and $H_{0}^{k+}$ for each $k$ separately. Clearly this requires that we proceed to stage 2 of  the adaptive trial unless all 4 stage 1 p-values are either above the futility bound $\alpha_{0}$ or below $\alpha_{1}$. A final judgement about bioequivalence, that is assessing $H_{0}^{1+}\cup H_{0}^{2+}\cup H_{0}^{1-}\cup H_{0}^{2-}$ can then be made using the intersection-union principle \cite{berger}. Specifically, with the traditional approach $H_{0}^{1-}\cup H_{0}^{2-}$ is rejected if 
  $\max_{k}\{Q(p_{1-}^{(k)},p_{2-}^{(k)},\alpha_{1},\alpha_{0})\}<\alpha$ and  $H_{0}^{1+}\cup H_{0}^{2+}$ is rejected if $\max_{k}\{Q(p_{1+}^{(k)},p_{2+}^{(k)},\alpha_{1},\alpha_{0})\}<\alpha$. In comparison our suggestion rejects $H_{0}^{1-}\cup H_{0}^{2-}$ if $Q(p^{min}_{1-},p^{min}_{2-},\alpha_{1},\alpha_{0})<\alpha$ and $H_{0}^{1+}\cup H_{0}^{2+}$ if  $Q(p^{max}_{1-},p^{max}_{2-},\alpha_{1},\alpha_{0})<\alpha$. By construction our suggestion is more powerful if $A_{1}=A_{2}$ since then $Q(p^{min}_{1-},p^{min}_{2-},\alpha_{1},\alpha_{0})\leq \max_{k}\{Q(p_{1-}^{(k)},p_{2-}^{(k)},\alpha_{1},\alpha_{0})\}$ and  $Q(p^{max}_{1-},p^{max}_{2-},\alpha_{1},\alpha_{0})\leq \max_{k}\{Q(p_{1+}^{(k)},p_{2+}^{(k)},\alpha_{1},\alpha_{0})\}$.

\section{Sample size reestimation}\label{samplesize}
The adaptive sample size algorithm proposed in \cite{maurer} is tailored to their proposed decision rule, which rejects $H_{0}$ at the end of stage 1 if both $H_{0}^{-}$ and $H_{0}^{+}$ are rejected.  Otherwise, based on their decision rule, the trial proceeds to stage 2, where both hypotheses, $H_{0}^{-}$ and $H_{0}^{+}$ are tested.  As previously stated, our decision rule differs, as it allows for decisions regarding $H_{0}^{-}$ and $H_{0}^{+}$ to be taken at different stages.  In this section, we develop a sample size reassessment algorithm specific to this proposed decision strategy.  

\vspace{5mm}
\noindent Similar to the SSR algorithm in \cite{maurer}, the stage 2 sample size, $N_{2}$ is selected in order to ensure an overall target power of $1-\beta$ for declaring bio-equivalence, under the alternative hypothesis corresponding to the interim estimates of $\theta$ and $\sigma$.  If we denote the estimated stage 1 power under this alternative hypothesis by,

$$
1-\beta_{1} = P_{\hat{\theta}_{1},\hat{\sigma}_{1}}(p_{1-} \leq \alpha_{1}, p_{1+} \leq \alpha_{1}),
$$

\noindent then the minimum required conditional power, $CP_{\hat{\theta}_{1},\hat{\sigma}_{1}}(p_{1-},p_{1+})$, in order to ensure an overall target power of $1-\beta$ for declaring bio-equivalence is given by,
 
$$
\frac{\beta_{1}-\beta}{\beta_{1}-\beta_{0}},
$$

\noindent where,

$$
1-\beta_{0} = P_{\hat{\theta}_{1},\hat{\sigma}_{1}}(p_{1-} < \alpha_{0},p_{1+} < \alpha_{0})
$$

\noindent is the probability the trial does not stop due to futility at stage 1, under the specified alternative.  The conditional power, $CP_{\hat{\theta}_{1},\hat{\sigma}_{1}}(p_{1-},p_{1+})$, will take different forms depending on the stage 1 outcome.  If at the end of stage 1, $p_{1-} < \alpha_{0}$ and $p_{1+} < \alpha_{0}$, then the conditional power will be given by one of the following four scenarios,

\[
CP_{\hat{\theta}_{1},\hat{\sigma}_{1}}(p_{1-},p_{1+})= 
\begin{dcases*}
1, & $p_{1-} \leq \alpha_{1}$ and $p_{1+} \leq \alpha_{1}$ \\
P_{\hat{\theta}_{1},\hat{\sigma}_{1}}(p_{2-} \leq A(p_{1-})|p_{1-}), & $\alpha_{1} < p_{1-} < \alpha_{0}$ and $p_{1+} \leq \alpha_{1}$ \\
P_{\hat{\theta}_{1},\hat{\sigma}_{1}}(p_{2+} \leq A(p_{1+})|p_{1+}), & $p_{1-} \leq \alpha_{1}$ and $\alpha_{1} < p_{1+} < \alpha_{0}$ \\
P_{\hat{\theta}_{1},\hat{\sigma}_{1}}(p_{2-} \leq A(p_{1-}),p_{2+} \leq A(p_{1+})| p_{1-},p_{1+}), & $\alpha_{1} < p_{1-} < \alpha_{0}$ and $\alpha_{1} < p_{1+} < \alpha_{0}$ \\
\end{dcases*}
\]

Here, $A(p)$ denotes the conditional error function associated with a specific combination function $C$, (see \cite{wassmer}, Sect. 6.3).  If we denote the desired stage 2 conditional power by $cp$, then under scenario 2, where $\alpha_{1} < p_{1-} < \alpha_{0}$ and $p_{1+} \leq \alpha_{1}$, the required stage 2 samples size per arm, assuming 1:1 randomization, is given by,

$$
N_{2} = \frac{2\hat{\sigma}_{1}^{2}(\Phi^{-1}(1-A(p_{1-}))-\Phi^{-1}(1-cp))^{2}}{(\Delta+\hat{\theta}_{1})^{2}},
$$

\noindent which follows the derivation given in \cite{wassmer}.  Similarly, for scenario 3, the required stage 2 sample size per arm would be given by,

$$
N_{2} = \frac{2\hat{\sigma}_{1}^{2}(\Phi^{-1}(1-A(p_{1-}))-\Phi^{-1}(1-cp))^{2}}{(\Delta-\hat{\theta}_{1})^{2}}
$$

\noindent For scenario 4, where both hypothesis are tested at the end of stage 2, the required sample size for stage 2 can be calculated based on the values of the conditional error functions calculated at the end of stage 1. The implementation of this approach may for instance be pursued in R based on functions from the R-packages powerTOST \cite{powerTOST} and mvtnorm \cite{mvtnorm}. The R-code of such an implementation is provided in chapter 6 of \cite{pattersonjones}. Alternatively a simulation based approach may be employed to obtain conditional powers for a given stage 2 sample size.  
\vspace{5mm}

\noindent \cite{wassmer} provides an in-depth overview of the potential issues that may arise for sample size re-estimation algorithms on based conditional power arguments using the interim parameter estimates. Briefly, the above formulas for the stage 2 sample size, are increasing functions in $p_{1}$. As the stage 1 p-value tends to 1, the required stage 2 sample size may balloon.  Additionally, values of $(\Delta + \hat{\theta}_{1})$ and $(\Delta - \hat{\theta}_{1})$ close to zero would also cause the required stage 2 sample size to balloon.  This may lead to designs with poor operating characteristics, e.g. large efficiency losses compared to standard trials design or group-sequential designs with similar expected sample sizes.  For relative stage two samples sizes, $N_{2}/N_{1}$, less than 4, \cite{wassmer} has demonstrated that the relative efficiency loss of the adaptive design compared to a standard design is minimal.  Such situations, where the relative stage 2 sample size exceeds 4 may be avoided by either specifying a maximum stage 2 sample size, or through judicious choice of the stage 1 futility bound, $\alpha_{0}$.     

\noindent So far we have neglected to address the case in which the p-value associated with one of the hypotheses, exceeds the futility bound at the interim analysis while the remaining p-value lies within the interval, $(\alpha_{1}, \alpha_{0})$.  In such an instance, one may benefit from increasing the precision of the estimated treatment difference and, or demonstrating non-inferiority, with respect to the hypothesis to be tested in stage 2.  

We will motivate the sample size re-estimation algorithm for this case in a similar vein to the above algorithm, in that we wish to select a stage 2 sample size, $N_{2}$, in order to ensure an overall target power of $1-\gamma$, under the alternative hypothesis corresponding to the interim estimates of $\theta$ and $\sigma$. Specifically the targeted power corresponds to the probability of rejecting exactly one hypothesis given that exactly one stage 1 p-value exceeds the futility bound, that is:

\begin{align*}
   1-\gamma=P_{\hat{\theta}_{1},\hat{\sigma}_{1}}(&\{p_{1-} < \alpha_{1}, p_{1+} \geq \alpha_{0}\}\cup\{p_{1-} \geq \alpha_{0}, p_{1+} < \alpha_{1}\}\cup\\
   &\{p_{1-} \in(\alpha_{1},\alpha_{0}), C(p_{1-},p_{2-})<c, p_{1+} \geq \alpha_{0}\}\cup\\ &\{p_{1+} \in(\alpha_{1},\alpha_{0}), C(p_{1+},p_{2+})<c, p_{1-} \geq \alpha_{0}\}|\\
   &\{p_{1-} < \alpha_{0}, p_{1+} \geq \alpha_{0}\}\cup\{p_{1-} \geq \alpha_{0}, p_{1+} < \alpha_{0}\}) 
\end{align*}

To relate the above power to the conditional power we also need to specify the following two probabilities,

\begin{align*}
1-\gamma_{1} &= P_{\hat{\theta}_{1},\hat{\sigma}_{1}}(p_{1-} < \alpha_{1}, p_{1+} \geq \alpha_{0}) + P_{\hat{\theta}_{1},\hat{\sigma}_{1}}(p_{1-} \geq \alpha_{0}, p_{1+} < \alpha_{1})\\
1-\gamma_{0} &= P_{\hat{\theta}_{1},\hat{\sigma}_{1}}(p_{1-} < \alpha_{0}, p_{1+} \geq \alpha_{0}) + P_{\hat{\theta}_{1},\hat{\sigma}_{1}}(p_{1-} \geq \alpha_{0}, p_{1+} < \alpha_{0}).\\
\end{align*}

 Here, $1-\gamma_{1}$ corresponds to the probability, under the specified alternative, that one hypothesis is rejected at the interim, while testing of the other hypothesis is stopped due to futility.  $1-\gamma_{0}$ then corresponds to the probability, under the specified alternative, that testing of one hypothesis is stopped due to futility at the interim, while the remaining hypothesis is not.  The probability under the specified alternative, of testing a single hypothesis in stage 2 when testing of the other hypothesis was stopped due to futility at the interim, is then given by, $\gamma_{1} - \gamma_{0}$.   

\vspace{5mm}
 
\noindent The overall target power under the specified alternative can then be expressed as,

$$
1-\gamma = \frac{1-\gamma_{1} +\iint_{S} CP_{\hat{\theta}_{1},\hat{\sigma}_{1}}(p_{1-},p_{1+})dP_{\hat{\theta}_{1},\hat{\sigma}_{1}}(p_{1-},p_{1+})}{1-\gamma_{0}},
$$

\noindent where, $S$ denotes the set, $\{p_{1-} \in(\alpha_{1},\alpha_{0}),p_{1+} \geq \alpha_{0}\} \cup \{p_{1-} \geq \alpha_{0}, p_{1+} \in (\alpha_{1},\alpha_{0})\}$.  The minimum required conditional power, $CP_{\hat{\theta}_{1},\hat{\sigma}_{1}}(p_{1-},p_{1+})$, required in order to ensure a target power of $1-\gamma$ to reject the one remaining hypothesis is given by,

$$
\frac{(1-\gamma)\cdot(1-\gamma_{0})-1+\gamma_{1}}{\gamma_{1}-\gamma_{0}}.
$$

\noindent In this setting, the conditional power may take one of two forms,

\[
CP_{\hat{\theta}_{1},\hat{\sigma}_{1}}(p_{1-},p_{1+})= 
\begin{dcases*}
P_{\hat{\theta}_{1},\hat{\sigma}_{1}}(p_{2-} \leq A(p_{1-})|p_{1-}), & $\alpha_{1} \leq p_{1-} < \alpha_{0}$ and $p_{1+} \geq \alpha_{0}$ \\
P_{\hat{\theta}_{1},\hat{\sigma}_{1}}(p_{2+} \leq A(p_{1+})|p_{1+}), & $p_{1-} \geq \alpha_{0}$ and $\alpha_{1} \leq p_{1+} < \alpha_{0}$ \\
\end{dcases*}
\]

\noindent Under the assumption of 1:1 randomization, the minimum stage two sample size, $N_{2}$, can be obtained using the formulas for the stage 2 sample size derived above.  The only difference being the value of the conditional power that is plugged into the sample size formulas.  For further details regarding the sample size procedures, please refer to the discussion in \cite{maurer}. We also refer to the excellent discussion in \cite{maurer} on how one may choose values of $\theta$ and $\sigma$ based on stage 1 data to get a more informed sample size re estimation.  

When two target parameters are assessed for bioequivalence simultaneously  as in Section \ref{multend} we, however, need to extend the standard sample size procedures slightly. Below we describe the main points of the extension. Our approach is simulation based and based on the assumption that $(\hat{\theta}_{j}^{(1)},\hat{\theta}_{j}^{(2)})$ follows a bivariate normal distribution (at least asymptotically) with mean $(\theta^{(1)},\theta^{(2)})$ and variance $\frac{1}{n_{j}}\Sigma$. The covariance matrix  $\Sigma$ is in turn defined by the within subject variance/covariance of  the $\log(C_{max})$ and $\log(AUC)$ measurements. For instance consider a two arm parallel or cross-over design with 1:1 randomization of treatments. In such a design let $\log(C_{max})_{il}$ and $\log(AUC)_{il}$ denote the measurements for the $i$th subject with the $l$th treatment $l=1,2$. With this notation  and assuming no missing measurements we have 
$$
\Sigma=A\cdot\Gamma\cdot A^{T},
$$
where $\Gamma=var\{\log(C_{max})_{i1},\log(AUC)_{il},\log(C_{max})_{i2},\log(AUC)_{i2}\}$ and 
$$
A=
\begin{bmatrix}
1&0&-1&0\\
0&1&0&-1
\end{bmatrix}.
$$
In a cross-over study the empirical variance matrix estimating $\Gamma$ follows a Wishart distribution with $n_{j}-1$ degrees of freedoom. For a two arm parallel study the empirical within treatment arm variance matrixes  estimating the $2x2$ block diagonal matrices of $\Gamma$ are independent and both follow wishart distributions with $n_{j}/2-1$ degrees of freedom. 

Consequently,  We may simulate realizations of $(\hat{\theta}_{j}^{(1)},\hat{\theta}_{j}^{(2)})$ based on their multivariate normal distribution. We may furthermore simulate realizations of $\hat{\Sigma}=A\cdot\hat{\Gamma}\cdot A^{T}$ by simulating $\hat{\Gamma}$ as described above. By combining these quantities as described in Section \ref{multend} we may produce realizations of  $Z_{j}^{min}, Z_{j}^{max}$ and in turn $p_{j-}^{min},p_{j+}^{max}$. 

The above recipe enables simulation based estimation of both stage 1 power probability:

$$
P_{\theta,\Gamma}(p_{1-}^{min}<\alpha_{1},\:p_{1+}^{max}<\alpha_{1})
$$
and conditional power probabilities

\begin{align*}
&cp^{1}_{\theta,\Gamma}(p_{1-}^{min})=P_{\theta,\Gamma}(p_{2-}^{min}<A(p_{1-}^{min})|p_{1-}^{min}),\\
&cp^{1}_{\theta,\Gamma}(p_{1+}^{max})=P_{\theta,\Gamma}(p_{2+}^{max}<A(p_{1+}^{max})|p_{1+}^{max}),\\
&cp^{2}_{\theta,\Gamma}(p_{1-}^{min},p_{1+}^{max})=P_{\theta,\Gamma}(p_{2-}^{min}<A(p_{1-}^{min}),\:\:p_{2+}^{max}<A(p_{1+}^{max})| p_{1-}^{min},p_{1+}^{max}),
\end{align*}

The fundamental difference between the sample size re estimation in  \cite{maurer} and the one proposed here lies in the different decision strategies. Specifically the probability of declaring bioequivalence  with their decision strategy in a given alternative $\theta,\sigma$ is given by

\begin{eqnarray*}
&&P_{\theta,\sigma}(p_{1-}<\alpha_{1},\:\:p_{1+}<\alpha_{1})\\
&&+\int_{0}^{\alpha_{1}}\int_{\alpha_{1}}^{\alpha_{0}}cp^{2}_{\theta,\sigma}(p_{1-},p_{1+})dP_{\theta,\sigma}(p_{1+},p_{1-})\\
&&+\int_{\alpha_{1}}^{\alpha_{0}}\int_{0}^{\alpha_{1}}cp^{2}_{\theta,\sigma}(p_{1-},p_{1+})dP_{\theta,\sigma}(p_{1+},p_{1-})\\
&&+\int_{\alpha_{1}}^{\alpha_{0}}\int_{\alpha_{1}}^{\alpha_{0}}cp^{2}_{\theta,\sigma}(p_{1-},p_{1+})dP_{\theta,\sigma}(p_{1+},p_{1-}).
\end{eqnarray*}
Since $cp^{2}_{\theta,\sigma}(p_{1-},p_{1+})$ is smaller than both $cp^{1}_{\theta,\sigma}(p_{1-})$ and 
$cp^{1}_{\theta,\sigma}(p_{1+})$ it is clear that the decision strategy in \cite{maurer} will yield less power to declare bioequivalence than the one outlined in this paper.  We elaborate on this in the simulation studies presented in the next section.

The sample size re-assessment when one stops for futility w.r.t one of the hypotheses already at stage 1  may seem irrelevant in the context of Section \ref{adaptost} where only one parameter is assessed for bioequivalence. Here one would expect that the probability of rejecting the other hypothesis at stage 1 is very high. Specifically, we noted in Section \ref{adaptost} that this probability is essentially 1 when the stage 1 sample is reasonably sized. However, in the context of Section \ref{multend} with multiple endpoints, where the null hypotheses target multiple parameters and are no longer disjoint, such a scenario may very well be envisaged  and cannot be discarded.

\section{Simulation performance}\label{Conc ex}
In this section we investigate the feasibility of our proposal for a two-stage adaptive design in the context of designing a parallel treatment arm trial for assessing the equivalence of two different modes of administration of a biologic for treatment of atopic dermatitis. The motivation for using an adaptive design such as the proposed is a lack of knowledge about the actual pharmaco-kinetic difference between the two modes of administration. Safe-guarding against this uncertainty up front in a traditional fixed design would lead to a very high sample-size. Also there is an up front acknowledged risk that the two modes of administration may not be bio-equivalent. If this is the case a built in futility stop is judged to be desirable.

The specific performance is considered for 
%the classical inverse normal combination test as well as 
for the robust combination test suggested by \cite{maurer}. We term this the  Maurer max combination test in what follows. %For each of these combination tests 
To judge performance we investigate the following features: 
\begin{itemize}
    \item[1] The ability of our procedure to achieve the sought power of declaring bioequivalence.
    \item[2] The coverage and general behavior of the suggested overall confidence interval.
\end{itemize}
These features are investigated both for one endpoint and two endpoints simultaneously. 

Finally a direct comparison between the suggested sample size re-estimation procedure and the sample size re-estimation procedure outlined in \cite{maurer} is made to gauge the consequence of the different decision strategies in terms of overall power.  

Before we proceed to simulations, however, we give a short introduction to the  Maurer max combination test with the purpose of bridging the exposition in \cite{maurer} to Section \ref{adaptost}.

\subsection{The Maurer max combination test}
Following the notation of \cite{maurer} we define $Z_{j}=\Phi^{-1}(1-p_j)$, where $p_{j}, j=1,2$ denote the stagewise p-values.  The max maurer combination test is then given by
$$
Z_{max}=max\{wZ_{1}+\sqrt{1-w^2}Z_{2},w_{\star}Z_{1}+\sqrt{1-w_{\star}^2}Z_{2}\},
$$ 
where $w$ and $w_{\star}$ are two prespecified weights between zero and one.
Assuming that $Z_{1}$ and $Z_{2}$ are independent and standard normal the p-value of the max combination test  is calculated as:

$$
P(Z_{max}>z_{max})=1-P(Z_{max}\leq z_{max})=1-P(wZ_{1}+\sqrt{1-w^2}Z_{2}\leq z_{max},\:\: w_{\star}Z_{1}+\sqrt{1-w_{\star}^2}Z_{2}\leq z_{max}).
$$

Accordingly, the combination function for the Mauer max combination test  is given by:
$$
C_{w,w_{\star}}(p_{1},p_{2})=1-F(\max\{\varepsilon(w,p_{1},p_{2}),\varepsilon(w_{\star},p_1,p_2)\})
$$
with
$$
\varepsilon(w,p_1,p_2)=w\Phi^{-1}(1-p_1)+\sqrt{1-w^{2}}\Phi^{1}(1-p_{2})
$$
and where for two standard normal-variables $\varepsilon_{w}$ and $\varepsilon_{w_{\star}}$ with correlation $w\cdot w_{\star}+\sqrt{(1-w^{2})(1-w_{\star}^{2})}$
$$
F(x)=P(\varepsilon_{w}\leq x, \:\: \varepsilon_{w_{\star}}\leq x). 
$$
Consequently, for another standard normal variable $\varepsilon_{1}$ with correlation $w$ and $w_{\star}$ to $\varepsilon_{w}$ and $\varepsilon_{w_{\star}}$, respectively, we may write the overall p-value as:

\begin{eqnarray*}
&&Q(p_1,p_2,\alpha_{1},\alpha_{0})=p_{1}I\{p_{1}\in (0,\alpha_{1})\cup(\alpha_{0},1)\}\\
&&+I\{p_{1}\in(\alpha_{1},\alpha_{0})\}[\alpha_{1}+P(Z_{1-\alpha_{0}}\leq \varepsilon_{1}\leq Z_{1-\alpha_{1}},\:\: \max\{\varepsilon_{w},\varepsilon_{w_{\star}}\}\geq \max\{\varepsilon(w,p_{1},p_{2}),\varepsilon(w_{\star},p_1,p_2)\})]
\end{eqnarray*}

Notice that for $w=w_{\star}$ the max Maurer combination test corresponds to the ordinary inverse normal combination test. 

%\subsection{The Inverse normal combination test}
%The inverse normal combination function is given by:
%$$
%C_{w}(p_{1},p_{2})=1-\Phi\{w\Phi^{-1}(1-p_{1})+\sqrt{1-w^{2}}\Phi^{-1}(1-p_{2})\}
%$$
%where $w$ is a prefixed weight between zero and one. 
%Accordingly we may rewrite $Q(p_1,p_2,\alpha_{1},\alpha_{0})$  in terms of bivariate normal probabilities %to 
%\begin{eqnarray*}
%&&Q(p_1,p_2,\alpha_{1},\alpha_{0})=p_{1}I\{p_{1}\in (0,\alpha_{1})\cup(\alpha_{0},1)\}\\
%&&+I\{p_{1}\in(\alpha_{1},\alpha_{0})\}\{\alpha_{1}+P(Z_{1-\alpha_{0}}\leq \varepsilon_{1}\leq %Z_{1-\alpha_{1}},\:\: \varepsilon_{w}\geq w\Phi^{-1}(1-p_{1})+\sqrt{1-w^{2}}\Phi^{-1}(1-p_{2}))
%\end{eqnarray*}
%where $\varepsilon_{1}$ and $\varepsilon_{w}$ are standard normal with correlation $w$. 

\subsection{Simulation setup and results}

In our simulation setup we consider a parallel arm study with a stage 1 sample size of 40 in each arm.
Initially we consider only one endpoint. Specifically, we simulate the $\log(C_{max})$ endpoint from a standard normal distribution with mean zero in the reference arm and mean $\theta=\log(0.87),\log(0.95),\log(1)$ in the test arm. For the residual error a standard deviation of 0.294 corresponding to a coefficient of variation of 0.3 is applied in both arms.  The bioequivalence threshold is set to $\Delta=\log(1.25)$.

 With an overall significance level $\alpha=0.05$ and a prespecified weights $w=\sqrt{0.5}$ and $w_{\star}=0.5$, $w_\star=\sqrt{0.5}$, or $w_\star=\sqrt{0.85}$ we consider the following combinations of efficacy and futility bounds: $(\alpha_{1},\alpha_{0})=(0.026,1), (0.028,0.5),(0.034,0.2)$ for $w_{\star}=0.5$,  $(\alpha_{1},\alpha_{0})=(0.030,1), (0.031,0.5), (0,034,0.2)$ for $w_{\star}=\sqrt{0.5}$, and $(\alpha_{1},\alpha_{0})=(0.029,1), (0.030,0.5), (0.033,0.5)$  for $w_{\star}=\sqrt{0.85}$. Sample size reassessment was made based on stage 1 ML parameter estimates as described in Section \ref{samplesize} to ensure an overall power of 0.9 and with a maximal stage 2 sample-size of 300.  Specifically this means that we are using $1-\beta=0.9$. Conditional powers are estimated for each considered stage 2 sample size by simulation with 10000 simulated two one-sided tests.     
 
 For each parameter configuration we simulate 5000 trials with sample size reassessment after stage 1 as described above if we proceed to stage 2. If we do not proceed to stage 2 the stage 2 sample size is set to zero. For each trial we calculate the 90\% confidence limits in accordance with  $\alpha=5\%$ as described in Section \ref{adaptost}.  The results are summarized in Table ~\ref{tab1}.
 
 \begin{table}
 \begin{tabular}{|ccc|ccc|ccc|}
 \hline
 $\theta$&$\alpha_1$&$\alpha_0$&upper CI<$\theta$& $\theta$< lower CI& upper CI$\leq$ lower CI & Avg. $n_{2}$ & \multicolumn{2}{c|}{Avg. power}\\
 \multicolumn{3}{|c|}{}&\multicolumn{3}{|c|}{}&&Overall & Stage 1\\
  \hline
\multicolumn{9}{|c|}{$w=\sqrt{0.5},\:\:w_{\star}=0.5$}\\
\hline
$\log(1)$ & 0.026 & 1&0.050&0.050&0&24.894&0.999&0.858\\
& 0.028 & 0.5 & 0.051  & 0.053&0  & 29.200&0.999&0.860\\
& 0.034 & 0.2 &0.051 & 0.053&0  &31.860&0.988&0.882\\
\hline
$\log(0.95)$ & 0.026 & 1&0.048&0.049&0&50.458&0.997&0.746\\
 & 0.028 &0.5 &0.049 & 0.055&0  &57.384&0.993&0.746\\
& 0.034 &0.2& 0.057 & 0.050&0 & 54.060&0.962&0.782 \\ 
\hline
$\log(0.87)$ & 0.026 & 1&0.053&0.051&0.0002&185.826&0.778&0.270\\
& 0.028 &0.5 & 0.056 & 0.048&0 & 168.371 & 0.761&0.273\\
&0.034&0.2&0.055&0.049&0 & 108.960 & 0.625&0.303\\
 \hline
\multicolumn{9}{|c|}{$w=w_{\star}=\sqrt{0.5}$}\\
\hline
$\log(1)$ & 0.026 & 1&0.050&0.049&0&22.526&1.000&0.872\\
& 0.028 & 0.5 & 0.053  & 0.053&0  & 29,308&0.999&0.858\\
& 0.034 & 0.2 &0.054 & 0.050&0  &32.160&0.991&0.884\\
\hline
$\log(0.95)$ & 0.026 & 1&0.053&0.051&0&48.253&0.996&0.757\\
 & 0.028 &0.5 &0.048 & 0.053&0  &56.739&0.994&0.744\\
& 0.034 &0.2& 0.051 & 0.048&0 & 54.300&0.964&0.783 \\ 
\hline
$\log(0.87)$ & 0.026 & 1&0.051&0.051&0.0004&184.724&0.745&0.272\\
& 0.028 &0.5 & 0.049 & 0.054&0 & 167.664 & 0.758&0.274\\
&0.034&0.2&0.046&0.049&0 & 110.940 & 0.622&0.293\\
 \hline
 \multicolumn{9}{|c|}{$w=\sqrt{0.5},\:\:w_{\star}=\sqrt{0.85}$}\\
\hline
$\log(1)$ & 0.029 & 1&0.049&0.044&0&20.473&1.000&0.875\\
& 0.030 & 0.5 &0.050& 0.055&0 & 26.287&0.999&0.865\\
& 0.033 & 0.2 &0.052 & 0.054&0  &31.260&0.988& 0.884\\
\hline
$\log(0.95)$ & 0.029 & 1&0.052&0.052&0&48.123&0.997&0.752\\
& 0.030 &0.5 &0.047 & 0.052&0  &53.079&0.991&0.754\\
& 0.033 &0.2& 0.052 & 0.050&0 & 56.160&0.962& 0.775\\
\hline
$\log(0.87)$ & 0.029 & 1&0.051&0.050&0.0004&182.060&0.753&0.276\\
& 0.030 &0.5 & 0.048 & 0.049&0 & 165.924 & 0.734&0.269\\
&0.033&0.2&0.047&0.052&0 & 111.480 & 0.620&0.292\\
 \hline
\end{tabular}
 \caption{ Summary of results. The sample performance of the 90\% CI is summarized by the fraction of 5000 simulations for which  upper CI <$\theta$,  $\theta$< lower CI, and upper CI$\leq$lower CI , respectively. The average stage 2  sample size  is denoted by Avg $n_2$. The sample fraction of times the  null hypothesis is rejected is denoted by Avg. power} \label{tab1}
 \end{table}
 
In optimistic scenarios lower futility seems to increase stage two sample size. for the pessimistic scenario ($\theta=\log(0.87)$) setting the futility according to $\alpha_{0}=0.2$ results in a notably  lower stage 2 sample size but also a notably lower overall power than for $\alpha_{0}=0.5$. For $\alpha_{0}=0.5$ stage 2 sample sizes are marginally higher compared  to $\alpha_{0}=1$ (no futility) for the very optimistic scenario ($\theta=\log(1)$, whereas for the pessimistic scenario they are lower. Overall powers are generally close to each other comparing $\alpha_{0}=0.5$ to $\alpha_{0}=1$. In terms of choice of $w_{\star}$ the higher values $\sqrt{0.5}.\sqrt{0.85}$  seem comparable and marginally better in terms of stage 2 sample size than $w_{\star}=0.5$. As expected the stage 1 power increases marginally when we decrease the futility.

Turning to the confidence intervals acceptable coverage probabilities close to  5\% are observed. When $\alpha_{0}=1$ Proposition \ref{thm1} no longer quarantees that the the lower bound is below the upper bound and in these scenarios we also observe instances with a lower bound above the upper bound. In accordance with Propostion \ref{thm1} this is not observed when $\alpha_{0}\leq0.5$.

Our second simulation study is a direct extension of the scenarios considered in the first simulation. Here we also include  the endpoint $\log(AUC)$. We simulate $\log(AUC)$ from the same marginal distribution as $\log(C_{max})$ and with a  correlation between these to measurements of 0.8. Othervise the setup in terms of effect size, $w, w_{\star}, \alpha_{0},\alpha_{1}, \alpha$, sample size reassessment, and number of simulations is identical to the first simulation study. Results are summarized in Table~\ref{tab2}.

 \begin{table}
 \begin{tabular}{|ccc|ccc|ccc|}
 \hline
 $(\theta^{(1)},\theta^{(2)})$&$\alpha_1$&$\alpha_0$&upper CI<$\theta$& $\theta$< lower CI& upper CI$\leq$ lower CI & Avg. $n_{2}$ & \multicolumn{2}{c|}{Avg. power}\\
 \multicolumn{3}{|c|}{}&\multicolumn{3}{|c|}{}&&Overall & Stage 1\\
  \hline
\multicolumn{9}{|c|}{$w=\sqrt{0.5},\:\:w_{\star}=0.5$}\\
\hline
$(\log(1),\log(1))$ & 0.026 & 1&0.028&0.025&0&40.328&1.000&0.784\\
& 0.028 & 0.5 &0.027 &0.027 & 0 &42.838 &0.998&0.803\\
& 0.034 & 0.2 &0.023 &0.021&0&47.580&0.986&0.828\\
\hline
$(\log(0.95),\log(0.95))$ & 0.026 & 1&0.024&0.026&0&73.430&0.995&0.653\\
 & 0.028 &0.5 &0.025 &0.028 & 0 &82.244&0.989&0.656\\
& 0.034 &0.2&0.026&0.027&0&74.160&0.944&0.697 \\ 
\hline
$(\log(0.87),\log(0.87))$ & 0.026 & 1&0.026&0.026&0&221.427&0.669&0.174\\
& 0.028 &0.5 &0.025&0.026&0&188.349&0.659&0.181\\
&0.034&0.2&0.024&0.022&0&112.740&0.507&0.199\\
 \hline
\multicolumn{9}{|c|}{$w=w_{\star}=\sqrt{0.5}$}\\
\hline
$(\log(1),\log(1))$ & 0.030 & 1&0.027&0.028&0&36.129&0.999&0.806\\
& 0.031 & 0.5 &0.024 &0.025 &0 &43.426 &0.999&0.810\\
& 0.034 & 0.2 &0.026&0.026 &0  &45.660&0.982&0.829\\
\hline
$(\log(0.95).\log(0.95))$ & 0.030 & 1&0.026&0.027&0&68.233&0.996&0.678\\
 & 0.031 &0.5 &0.026 &0.028  &0&77.211&0.990&0.680\\
& 0.034 &0.2&0.026  &0.027 &0 &74.580 &0.939&0.690 \\ 
\hline
$(\log(0.87),\log(0.87))$ & 0.030 & 1&0.023&0.027&0&215.724&0.637&0.191\\
& 0.031 &0.5 &0.027 &0.027 &0 & 186.185 &0.624 &0.188\\
&0.034&0.2&0.028&0.028&0 &108.720 &0.505 &0.205\\
 \hline
 \multicolumn{9}{|c|}{$w=\sqrt{0.5},\:\:w_{\star}=\sqrt{0.85}$}\\
\hline
$(\log(1),\log(1))$ & 0.029 & 1&0.026&0.023&0&32.727&1.000&0.816\\
& 0.030 & 0.5 &0.024&0.026 &0 &40.039&0.999&0.801\\
& 0.033 & 0.2 &0.027&0.026&0 &49.605&0.982&0.816\\
\hline
$(\log(0.95),\log(0.95))$ & 0.029 & 1&0.026&0.025&0&69.568&0.995&0.662\\
& 0.030 &0.5&0.027&0.023&0&74.771&0.993&0.676\\
& 0.033 &0.2&0.026&0.026&0 &75.476&0.941&0.690\\
\hline
$(\log(0.87),\log(0.87))$ & 0.029 & 1&0.022&0.025&0&217.447&0.633&0.182\\
& 0.030 &0.5 &0.028&0.030&0 &182.181&0.638&0.194\\
&0.033&0.2&0.025&0.025&0 &110.996&0.504&0.197\\
 \hline
\end{tabular}
 \caption{ Summary of results. The sample performance of the 90\% CI is summarized by the fraction of 5000 simulations for which  upper CI <$\theta$,  $\theta$< lower CI, and upper CI$\leq$lower CI , respectively.  The average stage 2  sample size  is denoted by Avg $n_2$. The sample fraction of times the null hypothesis is rejected is denoted by Avg. power} \label{tab2}
 \end{table}
 
The results of our second $C_{max}$ and $AUC$ endpoint simulation study largely mirror those obtained in the $C_{max}$ only endpoint simulation study with two major exceptions. 

Firstly, the coverage of our 90\% confidence interval is well above 90\% in all considered scenarios. This is in line with Proposition \ref{thm2} which ensures that the used stagewise shifted p-values in the true parameter values are only asymptotically p-clud instead of being asymptotically uniformly distributed. This guarantees that asymptotically the actual coverage is at least as large as the nominal coverage, but not that the actual coverage approaches the nominal coverage asymptotically.    

Secondly, We note that compared to the $C_{max}$ endpoint simulation study we need somewhat larger stage 2 sample sizes to achieve similar power to detect bioequivalence. Intuitively, this may be a bit surprising as the additional endpoint we have added is highly correlated to our initial endpoint with a correlation of 0.8, but it is in line with the observations made in \cite{pallmann}.

We investigate the impact of between endpoint correlation in more detail in a third simulation study.  In this simulation study we restrict ourselves to only studying performance when the true effect for both endpoints is $\theta^{(1)}=\theta^{(2)}=\log(1)$, when $\alpha_{0}=0.5$, and when $w=w_{\star}=0.5$. The standard deviation for $\log(C_{max})$ and $\log(AUC)$  is set to $0.294$.  Between endpoint correlations of 0.8, 0.9,0.95,0.99, and 1 are considered. Results of the simulations are given in Table \ref{tab3}.

 \begin{table}
 \begin{tabular}{|c|ccc|ccc|}
 \hline
 Correlation &upper CI<$\theta$& $\theta$< lower CI& upper CI$\leq$ lower CI & Avg. $n_{2}$ & \multicolumn{2}{c|}{Avg. power}\\
 \multicolumn{1}{|c|}{}&\multicolumn{3}{|c|}{}&&Overall & Stage 1\\
 \hline
0.80&0.024 &0.025 &0 &43.426 &0.999&0.810\\
0.90&0.032&0.028&0&36.834&0.999&0.830\\
0.95&0.038&0.035&0&35.190&0.999&0.841\\
0.99&0.045&0.048&0&31.247&0.999&0.854\\
1&0.055&0.053&0&28.228&0.999&0.870\\
\hline

\end{tabular}
 \caption{ Summary of results. The sample performance of the 90\% CI is summarized by the fraction of 5000 simulations for which  upper CI <$\theta$,  $\theta$< lower CI, and upper CI$\leq$lower CI , respectively. 
  The average stage 2  sample size  is denoted by Avg $n_2$. The sample fraction of times the null hypothesis is rejected is denoted by Avg. power} \label{tab3}
 \end{table}
 
 Table \ref{tab3} shows that stage 2 sample size decreases towards that of a single endpoint as correlation increases to 1. When the correlation is 1 results are as expected nearly identical to the single endpoint results in table \ref{tab1} for the scenario $\theta=\log(1),\: \alpha_{0}=0.5,\:w_{\star}=\sqrt{0.5}$. We also note that almost perfect correlation between the two endpoints is needed to achieve a performance similar to that of a single endpoint.

In a final simulation we investigate the magnitude of the expected gain in stage two sample size with the adaptive TOST procedure proposed in this paper compared to the adaptive procedure  proposed in \cite{maurer}. We consider a similar setup as in the first simulation study restricting ourselves to no stopping for futility ($\alpha_{0}=1$) for which the adaptive procedure in \cite{maurer} was originally developed. Results are summarized in table \ref{tab4}

 \begin{table}
 \begin{tabular}{|cc|cccc|cccc|}
 \hline
  \multicolumn{2}{|c|}{}&\multicolumn{4}{|c|}{Maurer et al}&\multicolumn{4}{|c|}{Adaptive TOST}\\
 $\alpha_1$&$\theta$& Avg. $n_{2}$ &Avg $n_{2} | n_{2}>0$ & \multicolumn{2}{c|}{Avg. power} & Avg. $n_{2}$&Avg $n_{2} | n_{2}>0$  & \multicolumn{2}{c|}{Avg. power}\\
 \multicolumn{2}{|c|}{}&&&Overall & Stage 1 &&&Overall & Stage 1\\
  \hline
\multicolumn{10}{|c|}{$w=\sqrt{0.5},\:\:w_{\star}=0.5$}\\
\hline
0.026 &$\log(1)$  &25.747&177.564&1.000&0.855&24.894&175.061&0.999&0.858\\
&$\log(0.95)$ &52.000&200.617&0.996&0.741&50.458&198.965&0.997&0.746\\
&$\log(0.87)$&191.112&256.252&0.780&0.254&185.826&254.417&0.778&0.270\\
\hline
\multicolumn{10}{|c|}{$w=w_{\star}=\sqrt{0.5}$}\\
\hline
0.030&$\log(1)$&22.474&182.123&1.000&0.877&22.526&175.981&1.000&0.872\\
&$\log(0.95)$&48.768&204.052&0.997&0.761&48.253&198.245&0.996&0.757\\
&$\log(0.87)$&187.101&258.641&0.743&0.277&184.724&253.602&0.745&0.272\\
 \hline
 \multicolumn{10}{|c|}{$w=\sqrt{0.5},\:\:w_{\star}=\sqrt{0.85}$}\\
\hline
0.029&$\log(1)$&23.615&170.627&0.999&0.862&20.473&164.311&1.000&0.875\\
&$\log(0.95)$&48.156&194.493&0.995&0.752&48.123&193.8890&0.997&0.752\\
&$\log(0.87)$&184.962&255.261&0.744&0.275&182.060&251.464&0.753&0.276\\
 \hline
\end{tabular}
 \caption{ Summary of results.  The average stage 2  sample size  is denoted by Avg $n_2$. The average stage 2 sample size among stage 2 sample sizes larger than zero is denoted by Avg $n_2|n_{2}>0$ The sample fraction of times the null hypothesis is rejected is denoted by Avg. power} \label{tab4}
 \end{table}
 
From table \ref{tab4} we note a minor but consistent gain in stage 2 sample size comparing the proposed adaptive TOST to the adaptive procedure in \cite{maurer}. This is in line with the theoretical comparison provided in Section \ref{samplesize} but clearly of little practical relevance in the scenarios considered in this simulation study.
 
\pagebreak

\section{Discussion}\label{discuss}
In this paper, we have presented a comprehensive two-stage adaptive TOST procedure for assessing bioequivalence. The approach, which incorporates formal stopping criteria, both in terms of futility and efficacy, at the interim analysis for both one-sided tests, has several advantages in comparison to previously published adaptive TOST procedures. 

First and foremost, through the specification of the overall shifted p-values, confidence intervals, consistent with the decision of whether to reject the null hypothesis of non-equivalence, can be generated. It is not readily apparent how to go about generating such consistent confidence intervals when both hypotheses are retested when the trial continues to the second stage, e.g. methods B, C and D of \cite{potvin}.

Methods B, C and D of \cite{potvin} relied on informal stopping criteria at the interim analysis, e.g.  prespecified thresholds for the observed power to demonstrate average bioequivalence. However, the observed power is directly related to the p-values associated with the stage 1 TOST procedure. Depending on the intent of the first stage design, i.e. pilot trial vs the sample size corresponding to a single stage study, the relevant threshold for the observed power may vary greatly and would necessitate the use of simulation to calibrate the threshold in order to ensure the proposed design has acceptable operating criteria. Our approach relies on the specification of formal stopping boundaries in terms of the stage 1  test statistics and hence the standard group sequential machinery can be applied when designing the trial. 

The proposed framework also includes a formal algorithm for continuing to the second stage in the situation that one of the test statistics for the TOST procedure fails to exceed the futility bound at the interim. The proposed SSR algorithm selects the second stage sample size in order to ensure an overall target power conditional on exactly one hypothesis failing to reach the futility bound at the interim analysis. In practice, the method is implemented by deriving a bound for the conditional power that ensures the overall target power is met on average.

An often-overlooked aspect in designing a clinical trial to demonstrate the average bioequivalence of two drug formulations, is the regulatory requirement to demonstrate average bioequivalence with respect to both Cmax and AUC.  It is common to power such a trial based on the requirements related to demonstrating average bioequivalence with regards to Cmax. In practice, this assumption is driven by the observed differences between the magnitudes of the variances for Cmax and AUC. This assumption is, however, not inherently true.  As a part of this paper, we have provided an efficient, simulation-based algorithm, to assess the operating characteristics of such designs. The algorithm derives its efficiency from simulating realizations from the joint distribution underlying the test statistics.

Our proposed implementation takes advantage of the pre-specified binding futility bound, by relaxing the threshold for rejecting the null hypothesis of non-equivalence. Additionally, crucial features of the proposed confidence intervals hinges on the use of a binding futility bound. However, based on the level of uncertainty at the design stage and/or the intent of the first stage of the design, e.g. pilot study, it may be advantageous to either forgo the inclusion of a futility bound or specify the use of a non-binding futility bound. For regulatory considerations regarding the use of binding/non-binding futility bounds, please refer to the current FDA guidance on adaptive trial designs, \cite{AdaptiveDesignFDA}. The flexibility of being able to continue to the second stage, while controlling the type 1 error rate in the strong sense, despite failing to exceed the futility bound at the interim may outweigh the benefits of incorporating a binding bound, including the ability to derive sensible CIs.  

We have chosen to use the combination function approach to control the type I error rate in the strong sense. The examples and simulations presented in the paper use the same combination functions for both one-sided hypothesis tests and Pocock type bounds for the efficacy stopping criteria at stage 1 and 2. These assumptions were driven by the design characteristics of the motivating example and ease of implementation. The theory presented is easily extended to accommodate combination functions and efficacy/futility bounds specific for each of the two hypotheses tested.  For any given application, it may be beneficial to explore the impact of various combination functions, as well as futility and efficacy bounds for the stagewise test statistics.    

To ensure the efficiency of our proposed sample size re-estimation algorithm based on the observed conditional power we introduced a limit on the maximum stage 2 sample size. A similar outcome could be achieved through a judicious choice of the stage 1 futility bound. On the other hand, sample size re-estimation algorithms using criteria other than the observed conditional power, e.g. conditional power based on a pre-specified minimal, clinically relevant effect size, or predictive power, either in a Bayesian or frequentist setting, have been proposed in the literature.  As there is no universally preferred method, the criteria underpinning the sample size re-estimation algorithm should be thoroughly assessed either analytically or via simulation. In terms of the motivating example presented in this paper, given the relatively large stage 1 sample size, we would expect the SSR algorithm based on the observed conditional power to perform well, in terms of the expected sample size. This is due to the precision associated with the parameter estimates at the interim analysis. Recently, \cite{Mehta} published an optimal promising zone sample size re-estimation algorithm when testing for superiority. The approach is based on assessing the observed conditional power to reject the null hypothesis at the end of the trial, assuming the effect size is equal to the minimal clinically relevant effect size. It may be interesting to explore whether such an approach can be applied to the equivalence setting, by basing the SSR algorithm on the maximum effect size considered to be clinically equivalent.

\bibliography{paper}

%\begin{appendices}

\section*{Appendix}

\subsection*{A.1.\enspace Proof of Proposition \ref{thm1}}

%\section{Proof of Proposition \ref{thm1}}
We divide the proof into 4 scenarios
\begin{itemize}
\item[]Scenario 1: $p_{1-}^{-\Delta}\in (0,\alpha_{1})\cup(\alpha_{0},1)$ and $p_{1+}^{-\Delta}\in (0,\alpha_{1})\cup(\alpha_{0},1)$
\item[] Scenario 2: $p_{1-}^{-\Delta}\in (\alpha_{1},\alpha_{0})$ and $p_{1+}^{-\Delta}\in (\alpha_{1},\alpha_{0})$
\item[] Scenario 3: $p_{1-}^{-\Delta}\in (0,\alpha_{1})\cup(\alpha_{0},1)$ and $p_{1+}^{-\Delta}\in (\alpha_{1},\alpha_{0})$
\item[] Scenario 4: $p_{1-}^{-\Delta}\in (\alpha_{1},\alpha_{0})$ and $p_{1+}^{-\Delta}\in (0,\alpha_{1})\cup(\alpha_{0},1)$
\end{itemize}

In Scenario 1 we have $p{-}^{\delta}=p_{1-}^{\delta}$ and $p_{+}^{\delta}=p_{1+}^{\delta}$. Consequently solving (\ref{lower}) yields $l=\hat{\theta}_{1}-\sigma_{1n}Z_{1-\alpha}$ and solving (\ref{upper}) yields $u=\hat{\theta}_{1}+\sigma_{1n}Z_{1-\alpha}$.  We conclude that $l<u$ in Scenario 1.
\\\\For scenario 2 we   note that 
$$
p_{-}^{\delta}\geq\alpha_{1}^{\delta}\textnormal{ and } p_{+}^{-\delta}\geq\alpha_{1}^{-\delta}
$$
and that for $\alpha<0.5$
$$
\alpha_{1}^{\delta}+\alpha_{1}^{-\delta}\geq I(Z_{1-\alpha_{1}}<\frac{\Delta}{\sigma_{1n}})+I(Z_{1-\alpha_{1}}\geq\frac{\Delta}{\sigma_{1n}})\cdot2\cdot\Phi(\frac{\Delta}{\sigma_{1n}}-Z_{1-\alpha_{1}})>2\cdot\alpha
$$
The first inequality above follows from minimizing $f(x,k)=\Phi(x-k)+\Phi(-x-k)$ as a function of x in the two scenarios $k>0$ and $k\leq0$. The second inequality follows from the assumption that $\frac{\Delta}{\sigma_{1n}}-Z_{1-\alpha_{1}}>-Z_{1-\alpha}$.\\\\This effectively means 
that since $p_{-}^{l}=\alpha$ we have
$$
p_{+}^{-l}>\alpha
$$
From this we conclude that $-u<-l$ in Scenario 2.
\\\\For Scenario 3 we first note that $p_{-}^{\delta}=p_{1-}^{\delta}$ so that $l=\hat{\theta}_{1}-\sigma_{1n}Z_{1-\alpha}$ We also note that the requirement $p_{1+}^{-\Delta}<\alpha_{0}$ entails that $\Delta-\hat{\theta}_{1}>0$. Furthermore, solving
$$
\alpha_{1}^{\delta}=\alpha
$$
yields
$$
\tilde{u}=\sigma_{1n}(Z_{1-\alpha_{1}}-Z_{1-\alpha})-\Delta.
$$
Now since $p_{+}^{\delta}>\alpha_{1}^{\delta}$ we deduce that $-u<\tilde{u}$. Consequently

$$
u-l>\Delta-\hat{\theta}_{1}+\sigma_{1n}(2\cdot Z_{1-\alpha}-Z_{1-\alpha_{1}})>0
$$
where the last inequality follows from our assumptions. \\\\Scenario 4 proceeds similarly to Scenario 3. We note that $u=\hat{\theta}_{1}+\sigma_{1n}Z_{1-\alpha}$ and that $\hat{\theta}_{1}+\Delta>0$ and finally that $l<\tilde{u}$, where $\tilde{u}$ is defined in Scenario 3. This entails that
$$
u-l>\hat{\theta}_{1}+\Delta+\sigma_{1n}(2\cdot Z_{1-\alpha}-Z_{1-\alpha_{1}})>0
$$
\begin{flushright}
$\qed$
\end{flushright}

\subsection*{A.2.\enspace Proof of Proposition \ref{thm1a}}

%\section{Proof of Proposition \ref{thm1}}
Similar to the proof of Proposition \ref{thm1} We consider 4 scenarios
\begin{itemize}
\item[]Scenario 1: $p_{1-}^{-\Delta,min}\in (0,\alpha_{1})\cup(\alpha_{0},1)$ and $p_{1+}^{-\Delta,max}\in (0,\alpha_{1})\cup(\alpha_{0},1)$
\item[] Scenario 2: $p_{1-}^{-\Delta,min}\in (\alpha_{1},\alpha_{0})$ and $p_{1+}^{-\Delta,max}\in (\alpha_{1},\alpha_{0})$
\item[] Scenario 3: $p_{1-}^{-\Delta,min}\in (0,\alpha_{1})\cup(\alpha_{0},1)$ and $p_{1+}^{-\Delta,max}\in (\alpha_{1},\alpha_{0})$
\item[] Scenario 4: $p_{1-}^{-\Delta,min}\in (\alpha_{1},\alpha_{0})$ and $p_{1+}^{-\Delta,max}\in (0,\alpha_{1})\cup(\alpha_{0},1)$
\end{itemize}
In the first scenario we have $p_{-}^{\delta,min}=p_{1-}^{\delta,min}$ and $p_{+}^{\delta,max}=p_{1+}^{\delta,max}$. As a consequence we get $l^{min}=\hat{\theta}_{1}^{min}-\sigma_{1n}^{min}Z_{1-\alpha}$ and $u^{max}=\hat{\theta}_{1}^{max}+\sigma_{1n}^{max}Z_{1-\alpha}$. Since $\hat{\theta}_{1}^{min}\leq\hat{\theta}_{1}^{max}$ it follows that $l^{min}<u^{max}$ in Scenario 1.
\\\\For Scenario 2 we note that 
$$
p_{-}^{\delta,min}\geq\alpha_{1}^{\delta,min}\textnormal{ and } p_{+}^{-\delta,max}\geq\alpha_{1}^{-\delta,max}
$$
and that for $\alpha<0.5$
\begin{align*}
\alpha_{1}^{\delta,min}+\alpha_{1}^{-\delta,max}&=\Phi(\delta/\sigma_{1n}^{min}-Z_{1-\alpha_{1}}+\Delta/\sigma_{1n}^{min})+\Phi(-\delta/\sigma_{1n}^{max}-Z_{1-\alpha_{1}}+\Delta/\sigma_{1n}^{max})\\
&\geq\min\Big\{1,\Phi(\frac{-\omega\kappa+\sqrt{\kappa^2+2\log(w)\frac{w-1}{w+1}}}{w-1})+\Phi(\frac{\kappa-\omega\sqrt{\kappa^2+2\log(w)\frac{w-1}{w+1}}}{w-1})\Big\}\\
&>2\alpha,
\end{align*}
where the first inequality follows from minimizing $f(x,k_{1},k_{2},\omega)=\Phi(x-k_{1})+\Phi(-\omega x-k_{2})$ as a function of $x$ and the second inequality follows directly from assumption (\ref{assump1}) in Proposition \ref{thm1a}. As in Proposition \ref{thm1} we conclude that $-u^{max}<-l^{min}$ in Scenario 2.  
\\\\For Scenario 3 we first note that $p_{-}^{\delta,min}=p_{1-}^{\delta,min}$ so that $l^{min}=\hat{\theta}_{1}^{min}-\sigma_{1n}^{min}Z_{1-\alpha}$ We also note that the requirement $p_{1+}^{-\Delta,max}<\alpha_{0}$ entails that $0<\Delta-\hat{\theta}_{1}^{max}\leq\Delta-\hat{\theta}_{1}^{min}$. Furthermore, solving
$$
\alpha_{1}^{\delta,max}=\alpha
$$
yields
$$
\tilde{u}=\sigma_{1n}^{max}(Z_{1-\alpha_{1}}-Z_{1-\alpha})-\Delta.
$$
Now since $p_{+}^{\delta,max}>\alpha_{1}^{\delta,max}$ we deduce that $-u^{max}<\tilde{u}$. Consequently

$$
u^{max}-l^{min}>\Delta-\hat{\theta}_{1}^{min}+(\sigma_{1n}^{min}+\sigma_{1n}^{max})Z_{1-\alpha}-\sigma_{1n}^{max}Z_{1-\alpha_{1}}>0
$$
where the last inequality follows from the assumption $(1+w)Z_{1-\alpha}-Z_{1-\alpha_{1}}>0$. \\\\Scenario 4 proceeds similarly to Scenario 3. We note that $u^{max}=\hat{\theta}_{1}^{max}+\sigma_{1n}^{max}Z_{1-\alpha}$ and that $0<\hat{\theta}_{1}^{min}+\Delta\leq \hat{\theta}_{1}^{max}+\Delta$. Furthermore, solving
$$
\alpha_{1}^{\delta,min}=\alpha
$$
yields
$$
\tilde{l}=\sigma_{1n}^{min}(Z_{1-\alpha_{1}}-Z_{1-\alpha})-\Delta.
$$
Now since $p_{-}^{\delta,min}>\alpha_{1}^{\delta,min}$ we deduce that $l^{min}<\tilde{l}$. Consequently

$$
u^{max}-l^{min}>\Delta+\hat{\theta}_{1}^{max}+(\sigma_{1n}^{min}+\sigma_{1n}^{max})Z_{1-\alpha}-\sigma_{1n}^{min}Z_{1-\alpha_{1}}>0
$$
where the last inequality follows from the assumption $(1+1/w)Z_{1-\alpha}-Z_{1-\alpha_{1}}>0$.

Finally to show that sufficient condition entails (\ref{assump1}) first note that 

$$
\rho=\frac{1}{2}(\frac{-\omega\kappa+\sqrt{\kappa^2+2\log(w)\frac{w-1}{w+1}}}{w-1}+\frac{\kappa-\omega\sqrt{\kappa^2+2\log(w)\frac{w-1}{w+1}}}{w-1})=\frac{-\kappa-\sqrt{\kappa^{2}+2\log(w)\frac{w-1}{w+1}}}{2}<0.
$$
Then for any $k>0$ we have $\Phi(\rho-k)+\Phi(\rho+k)\geq 2\Phi(\rho)>2\alpha$ from which (\ref{assump1}) follows

\begin{flushright}
$\qed$
\end{flushright}

\subsection*{A.3. \enspace Proof of Proposition \ref{thm2}}

First we state following lemma that is crucial in proving the statement.  The proof of the lemma follows from straightforward algebra.   
\begin{lemma}\label{upperlower}
Let $U,V$ be real numbers, put $P=I(U\leq V)$, and let $\sigma_{1},\sigma_{2}$ denote positive real numbers then the following holds:
\begin{align*}
    &\frac{U\cdot P+ V\cdot(1-P)}{P\cdot\sigma_{1}+(1-P)\cdot\sigma_{2}}\leq \frac{U}{\sigma_{1}}I(\sigma_{1}\leq\sigma_{2})+\frac{V}{\sigma_{2}}I(\sigma_{1}>\sigma_{2}),\\
    &\frac{U\cdot(1-P)+ V\cdot P}{(1-P)\cdot\sigma_{1}+P\cdot\sigma_{2}}\geq \frac{U}{\sigma_{1}}I(\sigma_{1}>\sigma_{2})+\frac{V}{\sigma_{2}}I(\sigma_{1}\leq\sigma_{2}).
\end{align*}
\end{lemma}

For proving Proposition \ref{thm2} first consider $\theta^{(1)}<\theta^{(2)}$. In this scenario we have 
$$
P(\hat{\theta}_{j}^{(1)}\geq\hat{\theta}_{j}^{(2)})\rightarrow 0 \text{ as } n_{j}\rightarrow\infty.
$$
This entails that 
$$
Z_{j}^{min}=\frac{\hat{\theta}_{j}^{(1)}+\Delta}{\sigma_{jn}^{(1)}}+o_{P}(1).
$$
Accordingly when $H_{0}^{1-}\cup H_{0}^{2-}\cap\{(\theta^{(1)},\theta^{(2)})|\:\:\theta^{(1)}<\theta^{(2)}\}=H_{0}^{1-}\cap\{(\theta^{(1)},\theta^{(2)})|\:\:\theta^{(1)}<\theta^{(2)}\}$ is true:

$$
\lim_{n_{j}\rightarrow\infty}P(p_{j-}^{min}\leq\alpha)=\lim_{n_{j}\rightarrow\infty}P(p_{j-}^{(1)}\leq\alpha)\leq\alpha.
$$
For the scenario $\theta^{(1)}>\theta^{(2)}$ we similarly get:
$$
Z_{j}^{min}=\frac{\hat{\theta}_{j}^{(2)}+\Delta}{\sigma_{jn}^{(2)}}+o_{P}(1).
$$
Accordingly  when $(H_{0}^{1-}\cup H_{0}^{2-})\cap\{(\theta^{(1)},\theta^{(2)})|\:\:\theta^{(1)}>\theta^{(2)}\}=H_{0}^{2-}\cap\{(\theta^{(1)},\theta^{(2)})|\:\:\theta^{(1)}>\theta^{(2)}\}$ is true:
$$
\lim_{n_{j}\rightarrow\infty}P(p_{j-}^{min}\leq\alpha)=\lim_{n_{j}\rightarrow\infty}P(p_{j-}^{(2)}\leq\alpha)\leq\alpha.
$$
Finally for the scenario $\theta^{(1)}=\theta^{(2)}$ we use Lemma \ref{upperlower} to see that
$$
Z_{j}^{min}\leq \frac{\hat{\theta}_{j}^{(1)}+\Delta}{\sigma_{jn}^{(1)}} I(\sigma_{jn}^{(1)}\leq\sigma_{jn}^{(2)}) +\frac{\hat{\theta}_{j}^{(2)}+\Delta}{\sigma_{jn}^{(2)}} I(\sigma_{jn}^{(1)}>\sigma_{jn}^{(2)})
$$
Now let $\sigma^{(k)}$ denote the limit in probability of $\sqrt{n_{j}}\sigma_{jn}^{(k)}$ for $k=1,2$ and $j=1,2$ and note that when $\sigma^{(1)}<\sigma^{(2)}$ we may use the above equality to conclude that:
$$
 \frac{\hat{\theta}_{j}^{(1)}+\Delta}{\sigma_{jn}^{(1)}} I(\sigma_{jn}^{(1)}\leq\sigma_{jn}^{(2)}) +\frac{\hat{\theta}_{j}^{(2)}+\Delta}{\sigma_{jn}^{(2)}} I(\sigma_{jn}^{(1)}>\sigma_{jn}^{(2)})=\frac{\hat{\theta}_{j}^{(1)}+\Delta}{\sigma_{jn}^{(1)}}+o_{p}(1).
$$
from which we conclude that when $(H_{0}^{1-}\cup H_{0}^{2-})\cap\{(\theta^{(1)},\theta^{(2)})|\:\:\theta^{(1)}=\theta^{(2)}\}=H_{0}^{1-}\cap\{(\theta^{(1)},\theta^{(2)})|\:\:\theta^{(1)}=\theta^{(2)}\}$ is true:

$$
\lim_{n_{j}\rightarrow\infty}P(p_{j-}^{min}\leq\alpha)\leq\lim_{n_{j}\rightarrow\infty}P(p_{j-}^{(1)}\leq\alpha)\leq\alpha.
$$
Similarly, when  $\sigma^{(1)}>\sigma^{(2)}$ we get that
$$
 \frac{\hat{\theta}_{j}^{(1)}+\Delta}{\sigma_{jn}^{(1)}} I(\sigma_{jn}^{(1)}\leq\sigma_{jn}^{(2)}) +\frac{\hat{\theta}_{j}^{(2)}+\Delta}{\sigma_{jn}^{(2)}} I(\sigma_{jn}^{(1)}>\sigma_{jn}^{(2)})=\frac{\hat{\theta}_{j}^{(2)}+\Delta}{\sigma_{jn}^{(2)}}+o_{p}(1).
$$
from which we also conclude that when $(H_{0}^{1-}\cup H_{0}^{2-})\cap\{(\theta^{(1)},\theta^{(2)})|\:\:\theta^{(1)}=\theta^{(2)}\}=H_{0}^{2-}\cap\{(\theta^{(1)},\theta^{(2)})|\:\:\theta^{(1)}=\theta^{(2)}\}$ is true:

$$
\lim_{n_{j}\rightarrow\infty}P(p_{j-}^{min}\leq\alpha)\leq\lim_{n_{j}\rightarrow\infty}P(p_{j-}^{(2)}\leq\alpha)\leq\alpha.
$$
Finally for the case $\sigma^{(1)}=\sigma^{(2)}$ note that: 
$$
Z_{j}^{min}\leq  \frac{\hat{\theta}_{j}^{(1)}+\Delta}{\min(\sigma_{jn}^{(1)},\sigma_{jn}^{(2)})}=\frac{\hat{\theta}_{j}^{(1)}-\theta^{(1)}}{\sigma_{jn}^{(1)}}+\frac{\theta^{(1)}+\Delta}{\min(\sigma_{jn}^{(1)},\sigma_{jn}^{(2)})}+o_{P}(1).$$
Thus when $(H_{0}^{1-}\cup H_{0}^{2-})\cap\{(\theta^{(1)},\theta^{(2)})|\:\:\theta^{(1)}=\theta^{(2)}\}$ is true, that is, when $\theta^{(1)}\leq\Delta$ it follows that:
$$
Z_{j}^{min}\leq \frac{\hat{\theta}_{j}^{(1)}-\theta^{(1)}}{\sigma_{jn}^{(1)}}+o_{P}(1)
$$
from which we conclude that
$$
\lim_{n_{j}\rightarrow\infty}P(p_{j-}^{min}\leq\alpha)\leq \lim_{n_{j}\rightarrow\infty} P(\frac{\hat{\theta}_{j}^{(1)}-\theta^{(1)}}{\sigma_{jn}^{(1)}}>Z_{1-\alpha})=\alpha.
$$
The proof of the second statement in Proposition \ref{thm2} follows along the same lines as above.

\begin{flushright}
$\qed$
\end{flushright}

%\end{appendices}

\end{document}